\pdfoutput=1
\documentclass[twocolumn,showpacs,showkeys,amsmath,amssymb,pra]{revtex4}

\usepackage[pdftex]{graphicx}
\usepackage{dcolumn}
\usepackage{bm}

\begin{document}

\title{Probability representation and quantumness tests\\for qudits and two-mode light states}

\author{Sergey N. Filippov$^1$}
\email{filippovsn@gmail.com}
\author{Vladimir I. Man'ko$^2$}%
\email{manko@sci.lebedev.ru}
\affiliation{%
$^1$ Moscow Institute of Physics and Technology (State University), Moscow Region 141700, Russia\\
$^2$ P. N. Lebedev Physical Institute of the Russian Academy of
Sciences, Moscow 119991, Russia }

\date{September 15, 2009}

\begin{abstract}
Using tomographic-probability representation of spin states,
quantum behavior of qudits is examined. For a general $j$-qudit
state we propose an explicit formula of quantumness witnetness
whose negative average value is incompatible with classical
statistical model. Probability representations of quantum and
classical $(2j+1)$-level systems are compared within the framework
of quantumness tests. Trough employing Jordan-Schwinger map the
method is extended to check quantumness of two-mode light states.
\end{abstract}

\pacs{03.65.Ta, 03.65.Sq, 03.65.Wj}

\keywords{quantumness test, quantumness witness, spin tomography,
qudit, two-mode light}

\maketitle

\section{\label{introduction} Introduction}
Quantumness and classicality of a system have been discussed
previously by many authors. The classical and quantum pictures of
physical phenomena are known to be quite different. For example,
the quantum behavior of a particle must satisfy some constraints
due to hierarchy of uncertainty relations which corresponds to
describing quantum states by nonnegative density operators. On the
other hand, the probability-distribution function of a classical
particle in phase space must be nonnegative while Wigner function
\cite{wigner}, quantum analogue of the classical probability
distribution, is permitted to get negative values. If one uses the
formal analogue of density operators for a classical particle, the
domain of such operators is permitted to contain Hermitian
trace-class nonpositive operators. From this viewpoint it is much
easier to think about differences between quantum and classical
pictures than to consider their similarities. Nevertheless, it is
of vital importance to find explicit criteria for distinguishing
classical and quantum properties, e.g., to relate quantumness of
two-qubit states with the existence of quantum correlations
expressed in terms of Bell (Bell-like) inequality violation
\cite{bell,chsh}. Besides, quantumness of a bipartite system can
be considered by means of entropy relations (see, e.g.,
\cite{li}). Here, we are not going to discuss these tests and the
attention is chiefly focused on the simple quantumness test for a
single system that has been proposed recently in
\cite{alicki-small,alicki-large,alicki-josephson,alicki-answer},
critically analyzed in \cite{zukowski}, and experimentally
realized for a two-level single system in
\cite{genovese-migdall,genovese-PRA-improved-test,genovese-JPhysConfSeries}.
The main idea of this test is to find observables $A$ and $B$ such
that the mean values $\langle A \rangle$, $\langle B \rangle$, and
$\langle B-A \rangle$ are greater than or equal to zero for all
possible states of a system. The classicality corresponds to
commutative algebra and if this is the case, the inequality
$\langle B^2 \rangle \ge \langle A^2 \rangle$ holds true for all
possible system states. Otherwise, if $\langle B^2 \rangle <
\langle A^2 \rangle$ for a particular state, then these
measurement results cannot be attributed to the standard classical
model and are to be treated as quantum ones. The observable
$B^2-A^2$ was called quantumness witness \cite{alicki-large}
because its negative average value for a given state is an
intermediate evidence that system behaves in quantum manner. It is
worth noting that the problem of distinguishing quantumness and
classicality is of vital importance for the practical realization
of quantum computer which is supposed to be essentially quantum to
work adequately. Though there is no conventional definition of
quantumness, we will use this concept in that sense that the
classical probabilistic model is incapable of accounting for
measurement outcomes.

Recently an appropriate method for considering similarity and
difference of classical and quantum features was suggested
\cite{tombesi-manko}, where the quantum states were associated
with the standard probability distributions like in classical
statistical mechanics. The approach called tomographic-probability
representation of quantum states was developed, e.g., in
\cite{filipp-08,filipp-qubit-portrait} and reviewed recently in
\cite{ibort}. This approach turned out to be convenient
\cite{filipp} to study quantumness test
\cite{alicki-small,alicki-large} given for qubits.

The aim of this paper is to interpret the test inequality
\cite{alicki-small,alicki-large} from the tomographic point of
view and propose quantumness witness for a general qudit state,
i.e., for particle with an arbitrary spin $j$. In regarding test
procedure we also indicate the difference between probabilistic
approaches to description of $(2j+1)$-level system, namely,
tomographic-probability representation and classical statistical
model. Another goal of this paper is to apply the quantumness test
to systems with continuous variables, for instance, two-mode light
states.

The paper is organized as follows.

In Sec. \ref{section:spin-tomogram}, we epitomize spin tomograms
and dual tomographic symbols. In Sec. \ref{section:test}, the test
inequality is developed by virtue of tomograms and explanation is
given why, using probabilities, the inequality can be violated in
quantum domain and is always valid in classical domain. In Sec.
\ref{section:witness}, an explicit formula of quantumness
witnetness is suggested for a general spin-$j$ case. In Sec.
\ref{section:two-mode-light}, quantumness of two-mode light states
is investigated by utilizing Jordan-Schwinger map to qudit states.
Finally, in Sec. \ref{section:conclusions}, conclusions are
presented.

\section{\label{section:spin-tomogram} Spin Tomograms and Tomographic Symbols of Operators}

In quantum mechanics, any state of a particle with spin $j$ can be
equivalently described either by $(2j+1)\times(2j+1)$ density
matrix $\rho$ or by the following probability-distribution
function called spin tomogram \cite{dodonovPLA,oman'ko-jetp}:

\begin{equation}
w_j(m,u) = \langle j m | u \rho u^{\dag} | j m \rangle = {\rm Tr}
\Big( \rho ~\hat{U}_j(m,u) \Big),
\end{equation}

\noindent where $| j m \rangle$ is an eigenstate of the angular
momentum operators $\hat{J}_z$ and $\hat{\bf{J}}^2$, $u$ is a
$(2j+1)\times(2j+1)$ unitary matrix of the irreducible
representation either of group $SU(2)$ or $SU(N)$ with $N=2j+1$,
and by $\hat{U}(m,u)$ we denote the so-called dequantizer operator
(as concerns star-product quantization schemes, the general
procedure of using this operator is discussed in
\cite{manko:star:2,omanko:vitale}). Any tomogram satisfies
normalization conditions of the form

\begin{eqnarray}
& \sum\limits_{m=-j}^{j} w_j(m,u) = 1, \\
& \displaystyle{\frac{2j+1}{8\pi^2}} \int \limits_{0}^{2\pi} d
\alpha \int \limits_{0}^{\pi} \sin\beta d \beta \int
\limits_{0}^{2\pi} d \gamma \ w_{j} (m,u(\alpha,\beta,\gamma)) =
1.
\end{eqnarray}

\noindent The latter equation implies that $u$ is a group element
of $SU(2)$ and is parameterized by Euler angles $\alpha$, $\beta$,
and $\gamma$.

Given the spin tomogram one can reconstruct the density operator
$\hat{\rho}$ by means of the special operator $\hat{D}_j(m,u)$
called quantizer (see, e.g.,
\cite{castanos,filipp-spin-tomography,filipp-chebyshev}). Namely,

\begin{eqnarray}
\hat{\rho} &=& \sum\limits_{m=-j}^{j} \frac{1}{8\pi^2}
\int\limits_{0}^{2\pi} d\alpha \int\limits_{0}^{\pi} \sin \beta
d\beta \nonumber\\
& & \times \int\limits_{0}^{2\pi} d\gamma ~
w_{j}(m,u(\alpha,\beta,\gamma)) \
\hat{D}_{j}(m,u(\alpha,\beta,\gamma)).
\end{eqnarray}

Dual tomographic symbol $w_j^d(m,u)$ of the operator $\hat{A}$ is
introduced by replacing $\hat{\rho}$ by $\hat{A}$ and substituting
$\hat{D}(m,u)$ for $\hat{U}(m,u)$ and vice versa
\cite{omanko:vitale,patr} and was anticipated in the paper
\cite{oman'ko-97}. To be precise,

\begin{equation}
w_A^d(m,u) = {\rm Tr} \Big( \hat{A}\hat{D}(m,u) \Big), \\
\end{equation}

\begin{eqnarray}
\hat{A} &=& \sum\limits_{m=-j}^{j} \frac{1}{8\pi^2}
\int\limits_{0}^{2\pi} d\alpha \int\limits_{0}^{\pi} \sin \beta
d\beta \nonumber\\
& & \times \int\limits_{0}^{2\pi} d\gamma ~
w_{A}^{d}(m,u(\alpha,\beta,\gamma)) \
\hat{U}_{j}(m,u(\alpha,\beta,\gamma)).
\end{eqnarray}

\noindent Dual tomographic symbols are particularly useful for
calculations of mean values of observables. Indeed, average values
can be expressed with the help of tomographic symbols only

\begin{eqnarray}
\label{eq:trace-by-tomograms} {\rm Tr} \Big( \hat{\rho} \hat{A}
\Big) = \sum\limits_{m=-j}^{j} \frac{1}{8\pi^2}
\int\limits_{0}^{2\pi} d\alpha \int\limits_{0}^{\pi} \sin \beta
d\beta \qquad\qquad\qquad && \nonumber\\
\times \int\limits_{0}^{2\pi} d\gamma ~
w_j(m,u(\alpha,\beta,\gamma)) \
w_{A}^{d}(m,u(\alpha,\beta,\gamma)). &&
\end{eqnarray}

\section{\label{section:test} Quantumness Tests in View of Tomograms}

Let us consider two possible probability descriptions of an
$N$-level system.

\subsection{Classical statistical model}
The state of a system is given by the point on the
$(N-1)$-simplex, i.e., by the probability vector
$(p_1,p_2,\dots,p_N)$ with $p_i \ge 0$, $\sum_{i=1}^{N} p_i = 1$.
Observable $A$ is considered as a function with possible outcomes
$A_1, A_2, \dots, A_N$ which correspond to appropriate system
states. Then the average value of $A$ reads

\begin{equation}
\langle A \rangle_{cl} = p_1 A_1 + p_2 A_2 + \dots + p_N A_N =
\sum\limits_{i=1}^{N} p_i A_i.
\end{equation}

\noindent Nonnegativity of $A$ for all possible states implies
that $A_i \ge 0$, $i=1,\dots,N$. Similarly, the condition $\langle
B \rangle _{cl} \ge \langle A \rangle _{cl}$ is followed by the
inequality $B_i \ge A_i$ for all $i=1,\dots,N$. If this is the
case, the mean value of $B^2-A^2$ is necessarily nonnegative. In
fact,

\begin{equation}
\langle B^2 \rangle _{cl} = \sum\limits_{i=1}^{N} p_i B_i^2 \ge
\sum\limits_{i=1}^{N} p_i A_i^2 = \langle A^2 \rangle _{cl}
\end{equation}

\noindent or in the matrix form

\begin{eqnarray}
\label{eq:inequality-class}
\langle B^2 \rangle _{cl} - \langle A^2 \rangle _{cl} \qquad\qquad\qquad\qquad\qquad\qquad\qquad\qquad ~~~ && \nonumber\\
= {\rm Tr} \left[ \left(%
\begin{array}{cc}
  p_1 & p_1 \\
  p_2 & p_2 \\
  \cdots & \cdots \\
  p_N & p_N \\
\end{array}%
\right)\left(%
\begin{array}{cccc}
  B_1^2 & B_2^2 & \cdots & B_N^2 \\
  -A_1^2 & -A_2^2 & \cdots & -A_N^2 \\
\end{array}%
\right) \right] \ge 0. ~~~ &&
\end{eqnarray}

\subsection{Tomographic probability representation}
It was shown in Sec. \ref{section:spin-tomogram}, that any state
of particle with spin $j$ is uniquely determined by its spin
tomogram. In a similar way, any operator $\hat{A}$ acting on
Hilbert space of states $| j m \rangle$ is equivalently described
by its ordinary or dual tomographic symbols. Suppose
$A_1,A_2,\dots,A_N$ are possible outcomes while measuring
$\hat{A}$, i.e., eigenvalues of $\hat{A}$. Then the average value
$\langle A \rangle_q = {\rm Tr} \left( \hat{\rho} \hat{A} \right)$
can be expressed in terms of tomogram $w_j(m,u)$ and spectrum
$\{A_i\}$. In the paper \cite{filipp}, the case of qubits
($j=1/2$) was considered in detail, where the authors used formula
(\ref{eq:trace-by-tomograms}) as well as the explicit form of
quantizer and dequantizer operators. Here we extend the formula
obtained to a general qudit case. Suppose $u_A^{\dag}$ is a
unitary matrix of group $SU(N)$ such that it reduces the
self-adjacent operator $\hat{A}$ to the diagonal form, i.e.,

\begin{equation}
\hat{A} = u_A^{\dag} \hat{A}_d u_A = u_A^{\dag} \left(%
\begin{array}{cccc}
  A_1 & 0 & \cdots & 0 \\
  0 & A_2 & \cdots & 0 \\
  \cdots & \cdots & \dots & \dots \\
  0 & 0 & \dots & A_N \\
\end{array}%
\right) u_A;
\end{equation}

\noindent then the average value of $\hat{A}$ reads

\begin{eqnarray}
{\rm Tr} \big( \hat{\rho} \hat{A} \big) = {\rm Tr} \left(
\hat{\rho} u_{A}^{\dag} \Big( \sum_{m=-j}^{j} A_m |jm\rangle
\langle jm| \Big) u_{A}\right)\qquad ~~~ && \nonumber\\
= \sum_{m=-j}^{j} A_m \langle jm| u_{A} \hat{\rho} u_{A}^{\dag}
|jm\rangle = \sum_{m=-j}^{j} w(m,u_A) A_m. ~~ &&
\end{eqnarray}

From this it follows that the average value of quantumness witness

\begin{eqnarray}
\label{eq:inequality-quant} \langle \hat{B}^2 \rangle_q - \langle
\hat{A}^2 \rangle_q  = {\rm Tr} \Bigg[  \left(%
\begin{array}{cc}
  w_j(1,u_B) & w_j(1,u_A) \\
  w_j(2,u_B) & w_j(2,u_A) \\
  \cdots & \cdots \\
  w_j(N,u_B) & w_j(N,u_A) \\
\end{array}%
\right) \qquad && \nonumber\\
 \times  \left(%
\begin{array}{cccc}
  B_1^2 & B_2^2 & \cdots & B_N^2 \\
  -A_1^2 & -A_2^2 & \cdots & -A_N^2 \\
\end{array}%
\right) \Bigg] ~&&
\end{eqnarray}

\noindent can be negative even if the inequality $\langle B
\rangle_q \ge \langle A \rangle_q \ge 0$ holds true for all
possible states.

Comparing formulas (\ref{eq:inequality-class}) and
(\ref{eq:inequality-quant}) one can see that both classical and
quantum behavior of a system are described by probabilities. The
difference is merely that the classical state is associated with
the set of constants $(p_1,p_2,\dots,p_N)$ while the quantum state
corresponds to the function $w_j(m,u)$ depending on an element $u$
of group $SU(N)$.

\section{\label{section:witness} Quantumness Witness for a General Qudit State}
The abundance of quantumness witnesses for an arbitrary qudit
state $\hat{\rho} \ne \openone/N$ was emphasized in
\cite{alicki-large}. In spite of this fact, the construction of
the quantumness witness for a given state $\hat{\rho}$ was not
presented in the explicit manner and relied much on an implicit
form of the quantumness witness for qubits. Here we suggest
explicit operators $\hat{A}$ and $\hat{B}$ that can fill this gap.

{\bf Proposition}. Given the diagonal $N\times N$ density matrix
$\hat{\rho}_d = {\rm diag}(r_1,r_2,\dots,r_N) \ne \openone/N$,
quantum behavior of this state can be checked by virtue of the
quantumness witness $\hat{B}_d^2 - \hat{A}_d^2$ with

\begin{eqnarray}
\label{eq:A&B-witness} \hat{A}_d = \| A_{ik}^{(d)} \| =
\frac{\Big( 1-a(r_i - \frac{1}{N}) \Big)^{1/2} \Big( 1-a(r_k -
\frac{1}{N}) \Big)^{1/2} }{\sum_{n=1}^{N}(r_n - \frac{1}{N})^2}, && \nonumber\\
\hat{B}_d = \hat{A}_d + b \hat{M}, \qquad \hat{M} = \| M_{ik} \| =
N \delta_{ik} - 1, \quad \quad ~ &&
\end{eqnarray}

\noindent where $a=\frac{3N}{4(N-1)}$, $b=\frac{1}{4(N-1)}$, and
$\delta_{ik}$ is the Kronecker symbol.

{\bf Proof}. It is evident that the operators $\hat{A}_d$ and
$\hat{B}_d$ are nonnegative by construction. Besides, the operator
$\hat{B}_d - \hat{A}_d = b \hat{M}$ has only nonnegative
eigenvalues so the requirement $\langle B_d \rangle \ge \langle
A_d \rangle \ge 0$ is met for all possible states. Let us now show
that ${\rm Tr} \left( \hat{\rho}_d (\hat{B}_d^2 - \hat{A}_d^2)
\right) < 0$.

Indeed, using explicit formulas of operators from the statement of
Proposition, we obtain

\begin{figure}
\includegraphics{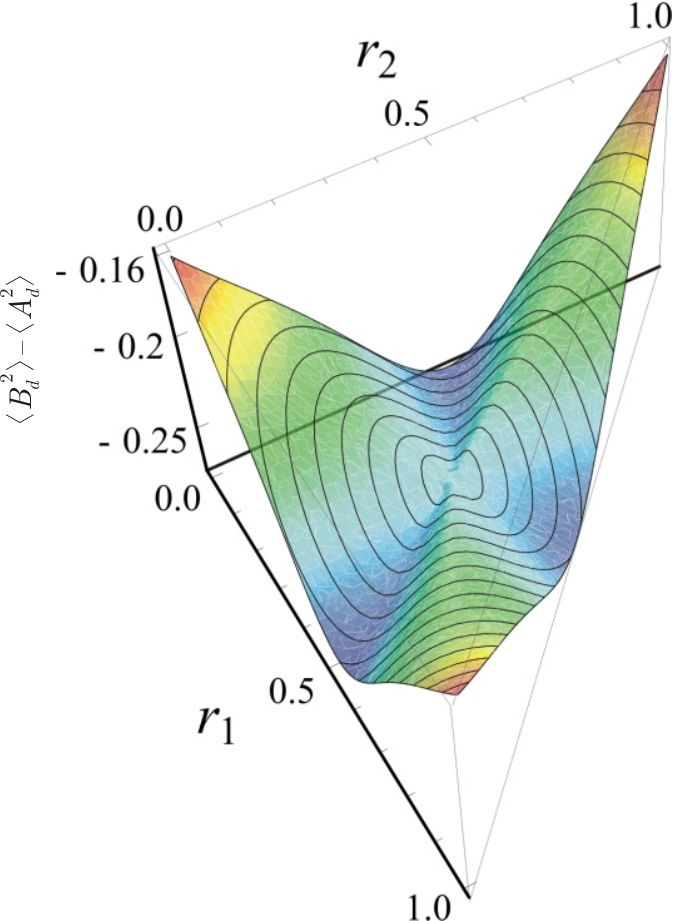}
\caption{\label{figure1} Quantum behavior of qutrit state
$\rho_d={\rm diag}(r_1,r_2,1-r_1-r_2)$ is pointed out by the
negativity of the average value of the quantumness witness
$\hat{B}_d^2-\hat{A}_d^2$ given by (\ref{eq:A&B-witness}). The
smaller the radius of the circle on the surface the closer one
approaches to the maximally mixed state $\rho_{cl}=(1/3,1/3,1/3)$.
Witness is not defined for this state because it is classical.}
\end{figure}

\begin{eqnarray}
\label{eq:witness-diag} {\rm Tr} \left( \hat{\rho}_d (\hat{B}_d^2
- \hat{A}_d^2) \right) = \frac{b}{{\rm Tr}\rho_d^2 -
\frac{1}{N}} \qquad\qquad\qquad\qquad\qquad && \nonumber\\
\times \Bigg\{ 2N \left[ 1 - a \Big({\rm Tr}\rho_d^2 -\frac{1}{N}
\Big) \right] + b N (N-1)
\Big({\rm Tr}\rho_d^2 -\frac{1}{N} \Big) && \nonumber\\
 - 2 \sum\limits_{k,l=1}^{N} r_k  \left( 1-a\Big(r_k -
\textstyle{\frac{1}{N}}\Big) \right)^{1/2} \left( 1-a\Big(r_l -
\textstyle{\frac{1}{N}}\Big) \right)^{1/2} \Bigg\}. && \nonumber\\
\end{eqnarray}

\noindent By $\Sigma$ denote the sum $\sum_{k,l=1}^{N}$ in the
equation above. The expression (\ref{eq:witness-diag}) takes its
maximal value when $\rho_d$ is a density matrix of pure state. If
this is the case,

\begin{equation}
\Sigma
> 1-a\Big(1 - \frac{1}{N}\Big) + (N-1) \left[ 1-a\Big(1 -
\frac{1}{N}\Big) \right]^{1/2}
\end{equation}

\noindent and consequently,

\begin{eqnarray}
\label{eq:x-sqrt[x]} {\rm Tr} \left( \hat{\rho}_d (\hat{B}_d^2
- \hat{A}_d^2) \right) \qquad\qquad\qquad\qquad\qquad\qquad\qquad && \nonumber\\
< 2 b N \Bigg\{ \left[ 1 - a \Big(1 -\textstyle{\frac{1}{N}} \Big)
\right] - \left[ 1 - a \Big(1 -\textstyle{\frac{1}{N}} \Big)
\right]^{1/2} \Bigg\} &&
\nonumber\\
+ b^2 N (N-1).~~ &&
\end{eqnarray}

\noindent The expression in braces at the right side of
(\ref{eq:x-sqrt[x]}) is less than zero whenever
$0<a<\frac{N}{N-1}$ and achieves minimal value $-1/4$ when
$a=\frac{3N}{4(N-1)}$. A proper choice of $b$, namely
$b=\frac{1}{4(N-1)}$, ensures

\begin{equation}
\label{eq:asymptote} {\rm Tr} \left( \hat{\rho}_d (\hat{B}_d^2 -
\hat{A}_d^2) \right) < - \frac{N}{16 (N-1)} < -\frac{1}{16} < 0,
\end{equation}

\noindent which is the best evidence of quantumness witness and
concludes the proof $\Box$.

\begin{figure}
\begin{center}
\includegraphics{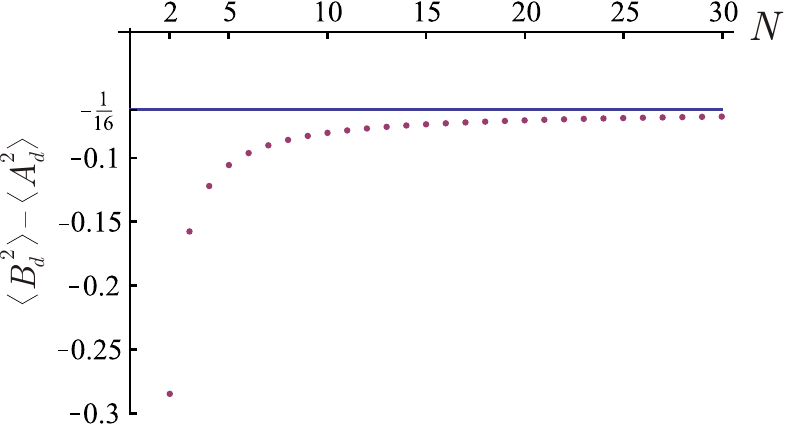}
\caption{\label{figure2} Maximum mean value of quantumness witness
against number of levels $N=2j+1$ (dots). Solid line is an
asymptote of this dependence and is predicted by
(\ref{eq:asymptote}). Negativity of all the values allows to
detect quantumness of any qudit state with an arbirary spin $j$.}
\end{center}
\end{figure}

To exemplify this Proposition we consider the case of qutrits
($j=1$, $N=3$) and plot the corresponding average value of
quantumness witness versus parameters $r_1$ and $r_2$ of the
diagonal density matrix $\rho_d$ (see Fig. \ref{figure1}). The
pattern in Fig. \ref{figure2} confirms that the proved Proposition
is applicable not only to qutrits but also to an arbitrary spin
system.

Proposition is followed by a simple

{\bf Consequence}. Let $u$ be a unitary matrix of group $SU(N)$
that reduces the given density matrix $\rho$ to the diagonal form,
i.e., $\rho = u \rho_d u^{\dag}$. Then the operator
$\hat{B}^2-\hat{A}^2$ with $\hat{A}=u \hat{A}_d u^{\dag}$ and
$\hat{B}=u \hat{B}_d u^{\dag}$ is a quantumness witness for the
state $\rho$.

\section{\label{section:two-mode-light} Quantumness of Two-Mode Light States}
Applying the Jordan-Schwinger map \cite{jordan,schwinger} to
qudits, one can readily extend quantumness tests to two-mode light
states. By construction, the following correspondence is
ascertained:


\begin{center}
\begin{tabular}{c|c}
  {\rm qudit with spin $j$} & {\rm two oscillators} \\
  \hline
  $\openone$ & $\openone$ \\
  $\hat{J}_{+}$ & $\hat{a}^{\dag}\hat{b}$ \\
  $\hat{J}_{-}$ & $\hat{a}\hat{b}^{\dag}$ \\
  $\hat{J}_z$ & $\frac{1}{2} ( \hat{a}^{\dag}\hat{a} -
\hat{b}^{\dag}\hat{b} )$ \\
  $|jm\rangle$ & $|n_a,n_b\rangle \ {\rm with} \  m=\frac{n_a-n_b}{2},\
j=\frac{n_a+n_b}{2}$ \\
\end{tabular}
\end{center}

\noindent Here, $\hat{a}$ and $\hat{a}^{\dag}$ ($\hat{b}$ and
$\hat{b}^{\dag}$) are annihilation and creation operators in each
mode, $n_a$ and $n_b$ are numbers of photons in the first and the
second modes, respectively. This analogy implies that in order to
check the quantumness of the state $|n_a n_b\rangle$ one should
follow the procedure:

(i) map this state onto qudit state
$|jm\rangle$ with $j=(n_a+n_b)/2$ and $m=(n_a-n_b)/2$;

(ii) construct quantumness witness $\hat{B}^2-\hat{A}^2$ for state
$|jm\rangle$ in accordance with the Proposition in the previous
section;

(iii) using the table above, fulfil inverse mapping of the
constructed witness onto operator acting on Hilbert space of
two-mode light states;

(iv) make sure that the average value of the final witness is
negative for the given state $|n_a n_b\rangle$.

The remarkable fact is that the proposed approach is incapable to
indicate the quantumness of the vacuum state. In this sense the
method developed is analogues to quasiprobabilistic treatment of
Wigner function \cite{wigner}.

\section{\label{section:conclusions}Conclusions}

To summarize we underline main results of the paper. The article
is concerned with $N$-level systems (qudits with spin $j$,
$N=2j+1$) and then is extended to deal with two-mode light states.
We managed to consider quantumness tests proposed recently from
the probabilistic point of view. According to this approach, we
regarded both classical and quantum states by virtue of
probabilities, the former being described by constant numbers and
the latter being associated with the probability-distribution
function depending on a group element. This difference between two
descriptions explains the reason why the test inequality is always
fulfilled in classical domain and can be violated in quantum one.
The explicit form of quantumness witness is suggested for a
general qudit case. Apart from the analytical proof, some examples
are also given to check an adequate work and illustrate the
usefulness of this witness. Jordan-Schwinger map of spin states
onto two-mode light states is utilized as a key step toward
quantumness tests for systems with continuous variables. We hope
to extend quantumness witness formulated in the work for second
momenta to the inequalities containing higher moments.

\begin{acknowledgments}
V.I.M. thanks the Russian Foundation for Basic Research for
partial support under Project Nos. 07-02-00598 and 08-02-90300.
S.N.F. thanks the Ministry of Education and Science of the Russian
Federation and the Federal Education Agency for support under
Project No. 2.1.1/5909. The authors are grateful to the Organizers
of the Eleventh International Conference on Squeezed States and
Uncertainty Relations (Olomouc, Czech Republic, June 22-26, 2009)
for invitation and kind hospitality. S.N.F. would like to express
his gratitude to the Organizing Committee of the Conference and
especially to Professor Jan Perina Jr. for financial support.
\end{acknowledgments}

\end{document}